\documentstyle[12pt,twoside]{article}
\def\gsim{\mathrel{\rlap{\lower4pt\hbox{\hskip1pt$\sim$}}
    \raise1pt\hbox{$>$}}}         
\newcommand{\shat}{\hat s}

\newlength{\dinwidth}
\newlength{\dinmargin}
  \newenvironment{defl}[1]%
  {\begin{list}{}{\settowidth{\labelwidth}{#1}%
  \setlength{\leftmargin}{\labelwidth}%
  \addtolength{\leftmargin}{\labelsep}%
  \setlength{\itemsep}{0pt plus 1pt}
  \setlength{\parsep}{0pt plus 1pt}
  \setlength{\topsep}{0pt plus 1pt}
  \setlength{\partopsep}{0pt plus 1pt}
  \setlength{\parskip}{2mm plus 1mm minus 1mm}
  }}%
  {\end{list}}
\begin{document}
\thispagestyle{empty}   
\noindent
DESY 96--058         \hfill ISSN 0418-9833\\
April 1996       \\
\begin{center}
  \begin{Large}
  \begin{bf}
AROMA 2.2 -- A Monte Carlo Generator\\
for Heavy Flavour Events in $ep$ Collisions\\
  \end{bf}
  \end{Large}
  \vspace{5mm}
  \begin{large}
G.\ Ingelman$^{a,b}$, J.\ Rathsman$^{b}$, 
G.A.\ Schuler\footnote{Heisenberg fellow}$^{c}$\\
  \end{large}
  \vspace{3mm}
ingelman@desy.de ~~~~ rathsman@tsl.uu.se ~~~~ schulerg@afsmail.cern.ch \\
\end{center}
\vspace{3mm}
$^a$ Deutsches~Elektronen-Synchrotron~DESY,
Notkestrasse~85,~D-22603~Hamburg,~FRG\\
$^b$ Dept. of Radiation Sciences, Uppsala University,
Box 535, S-751 21 Uppsala, Sweden\\
$^c$ Theory Division, CERN, CH-1211 Geneva 23, Switzerland
%
\begin{quotation}
\noindent
{\bf Abstract:}
A program to simulate the production of heavy quarks through
the boson-gluon fusion process in $e^{\pm}p$ collisions is presented.
The full electro\-weak structure of the electron--gluon
interaction is taken into account as well as the masses of
the produced heavy quarks. Higher order QCD radiation is treated 
using initial and final state parton showers, and hadronization is
performed using the Lund string model. 
Physics and programming aspects are described in this manual.
\end{quotation}
%
\section{Physics of included processes}
The production of heavy quarks in $ep$ collisions is dominated by
boson-gluon fusion (BGF) into a heavy quark-antiquark pair
\begin{equation}
\label{BGQQ}
  V(q) + g(p) \rightarrow Q_f(p_f) + \bar{Q}_{f'}(p_{f'})
\ ,
\end{equation}
occuring as ${\cal O}(\alpha_s \alpha^2)$
parton level subprocess in the $ep$ scattering process
\begin{equation}
\label{EPQQ}
  e^{\pm}(l_e) + p(P) \rightarrow l'(l') + Q_f(p_f)
                       + \bar{Q}_{f'}(p_{f'}) + X
\ .
\end{equation}
Here the symbols in brackets denote the corresponding four-momenta.
In the charged current (CC)
case $V$ is the $W^{\pm}$-boson whereas in the neutral current (NC)
case it corresponds to $\gamma/Z^0$ exchange and the produced
quark and antiquark have the same flavour $f$.
Beside the normal DIS variables defined by
\begin{equation}
\label{DISVAR}
  Q^2   \equiv   -q^2 = -(l_e - l')^2
   \quad , \quad
  W^2   \equiv   (P+q)^2
   \quad , \quad
  x   \equiv   \frac{Q^2}{2P\cdot q}
   \quad , \quad
  y   \equiv   \frac{P\cdot q}{P\cdot l_e}
\ ,
\end{equation}
out of which only two are independent, three additional independent
variables are needed to completely specify the process in eq.
(\ref{EPQQ}).
These are here taken as the momentum fraction $x_g$ of the gluon
relative to the proton, i.e.\ $p=x_gP$, the variable
\begin{equation}
\label{ZDEF}
   z = \frac{P\cdot p_f}{P\cdot q}
\ ,
\end{equation}
and the azimuthal (around the boson axis) angle $\Phi$
between the lepton and hadron planes
\begin{equation}
\label{PHIDEF}
  \cos\Phi = \frac{(\vec{p}\times\vec{l_e})\cdot(\vec{p}
     \times\vec{p_f})}
                   {\mid\vec{p}\times\vec{l_e}\mid\mid\vec{p}
     \times\vec{p_f}\mid}
\end{equation}
measured in the boson-gluon CM frame, i.e. $\vec{p}_f+\vec{p}_{f'}=0$.
The gluon momentum fraction is related to the usual Bjorken-$x$
variable by
\begin{equation}
\label{XGXREL}
   x_g = x + \frac{\shat}{ys} \geq x
\ .
\end{equation}
The variable $z$ is related to the angle of the
$Q\bar{Q}'$-axis with respect to the boson-gluon axis in this subsystem
cms and gives the heavy quark transverse momentum through the relation
$p_{\perp}^2=\hat{s}z(1-z)+z(m_{f'}^2-m_f^2)-m_f^2$.
As usual, $\shat$ is taken to denote the invariant mass square of the
$Q\bar{Q}'$-subsystem, i.e. $\shat = ( p_f + p_{f'} )^2$.

The cross section for process (\ref{EPQQ}) is then given by
\begin{equation}
\label{SIGMA}
  \sigma(e^{\pm}p\rightarrow Q\bar{Q}'X) =
      \int dy\,   \int dQ^2\,  \int dx_g\,  \int dz\,  \int d\Phi \;
         g(x_g,M_g^2) \, h(y,Q^2,x_g,z,\Phi)
\end{equation}
which is a convolution
of the gluon density $g(x_g,M_g^2)$ and a QCD part $h$ for the
subprocess. The latter
has been calculated in ref. \cite{GS} taking proper account
of the heavy quark masses and the complete electroweak structure for
both charged and neutral current processes including the
$\gamma - Z^0$ interference and allowing for arbitrary longitudinally
polarization of the $e^{\pm}$ beam. Both the deep inelastic and
photoproduction ($Q^2\to 0$) region is covered.

For the numerical evaluation \cite{GS,IS} of this cross section
the electroweak parameters are taken from {\sc Lepto} 6.5 \cite{LEPTO}.
The Kobayashi-Maskawa matrix elements have the default values of
{\sc Lepto} 6.5 and the quark masses the default values of {\sc Jetset} 7.4
\cite{JETSET}, in particular $m_c=1.35\,$GeV and $m_b=5\,$GeV.
All these default values can be changed by the user.
The mass scales for the gluon density and for $\alpha_s$,
$M_g$ and $M_s$, are taken to be $M_g = M_s = \sqrt{\hat{s}}$;
the number of flavours, $N_f$, and $\Lambda_{QCD}$
used in the first order expression for $\alpha_s$
are taken from {\sc Jetset} which means that $\Lambda_{QCD}$ is taken
from the parton densities.
A selection of parametrizations for the gluon density in the proton
are provided as alternatives (cf.\ LST(15) \cite{LEPTO}).
The numerical uncertainties in the cross sections are discussed in
\cite{IS}.

The next-to-leading order (NLO) QCD corrections to the inclusive heavy
quark distributions (rapidity and $p_\bot$) have been calculated for
photoproduction ($Q^2 \ll \Lambda^2$) \cite{NLOPHOTO} and
recently also for deep-inelastic scattering (DIS), i.e.\
$Q^2 \gg \Lambda^2$ \cite{NLOEP}. In the case of photoproduction,
the ${\cal O}(\alpha_s^2)$ corrected distributions
are found to be similar in shape to the leading-order ones, i.e.\
they can be described by an approximately constant $K$-factor.
It is therefore expected that the distributions from {\sc Aroma} describe the
shapes of inclusive heavy quark distributions
reasonably well, whereas the absolute normalization is
underestimated \cite{NLOPHOTO}. In contrast, the $K$-factor of DIS
is strongly kinematics dependent and therefore
not easily taken into account. In addition, the available NLO result
only applies for inclusive quantities and not for the exclusive final state
needed for MC simulation.

For a proper description of the complete
event topology higher order corrections
are important. Gluon bremsstrahlung
off the heavy quark and antiquark can be approximately taken into
account by the use of parton shower (PS) simulation algorithms.
From the experience of jets in $e^+e^-$ annihilation
one expects this approach, which includes the leading logarithm
contribution of higher corrections, to provide a better description of
detailed jet properties, such as hardness and width, as compared to
exact next-to-leading order matrix elements.
Such a shower algorithm \cite{JETSET} is therefore applied, but one
should note that the rate of clearly separable additional jets
need not be quite correct.

Production of heavy quarks in lowest order, eq. (1), is characterized by
a pair of heavy quarks that are essentially back-to-back
in the boson--proton CMS and with small overall transverse momentum
$p_\bot ({Q\bar{Q}})$. Small deviations from this can
arise from a non-vanishing primordial $k_\bot$ of
the incoming gluon, which is here generated according to a Gaussian
distribution.
Larger deviations from this topology may arise through parton radiation
from the gluon entering the fusion process.
Such initial state radiation also affects
the inclusive heavy quark spectra, which in NLO accuracy follow
rather closely those at the Born level (at least at low $Q^2$).
This implies constraints on the initial radiation and therefore one does
not expect large effects. Initial state parton showers are
implemented by suitably modifying the algorithm of ref.\ \cite{BS},
in particular allowing for a smooth $Q^2 \to 0$ limit.

In $ep$ scattering heavy quarks can also be produced through
{\em (i)} mixing in the leading order charged current process $e+q\to
\nu+Q$ \cite{IS}, {\em (ii)} the scattering on intrinsic charm
quarks in the proton \cite{IC}, {\em (iii)} resolved photon
contributions leading to hadronic production processes of heavy
flavours (e.g. $gg \to Q\bar{Q}$) and {\em (iv)} diffractive heavy
flavour excitation \cite{POMHAD}.
These processes, which are not included in {\sc Aroma}, are expected to have
much smaller cross sections than the BGF process in the HERA energy
range.

The {\sc Aroma} program is also applicable to the case of light flavours.
However, without the heavy quark mass as a cutoff against the divergences
in the matrix element it is then necessary to apply a lower cut on
the $p_\bot$ of the produced quarks in the boson-gluon fusion cms.
The program thus allows to set a $p_\bot$-cutoff, which may also be applied 
in heavy quark generation. 

\section{The Monte Carlo implementation}
The Monte Carlo
simulation model is based on the following main ingredients:
{\it (i)}
The complete matrix elements \cite{GS} to order $\alpha^2 \alpha_s$
for the boson-gluon fusion process, eq.\ (\ref{SIGMA}),
(including the masses of the produced heavy quarks and the full
electroweak structure of the interaction),
{\it (ii)}
gluon emission from the $Q\bar{Q}'$ system in a parton shower approach
\cite{JETSET},
{\it (iii)} initial state parton shower,
{\it (iv)} possible soft colour interactions as a mechanism for 
rapidity gaps \cite{SCI},
{\it (v)} hadronization with the Lund string model
\cite{LUND,JETSET} and heavy flavour decay.

The importance sampling method in {\sc Divonne} \cite{DIVON} is used to
generate phase space points according to the differential cross-section
formula in eq.~(\ref{SIGMA}). From the phase
space point ($y, Q^2, x_g, z, \Phi$), the four-momenta
of all partons (particles) are calculated.

Similarly to the $q\bar{q}$ pair produced in $e^+e^-$ annihilation the
heavy $Q\bar{Q}'$ pair can emit bremsstrahlung gluons (some of which may
split perturbatively into gluon or $q\bar{q}$ pairs) thereby creating a
cascade, or shower, at the parton level. To simulate this the
model for $e^+e^-$ \cite{JETSET} is applied.
In doing so an uncertainty arises as to what scale to use for the
maximum off-shellness of
the heavy quark and antiquark that initiate the
shower. Different options are provided, see LSTHFL(7), with the
default chosen as $(m_{\perp Q} + m_{\perp \bar{Q}'})^2$ since this
measures the momentum transfer in similarity with the case in the
$gg\to Q\bar{Q}$ process in hadron collisions.
To account for parton radiation from the incoming gluon, the initial
state parton shower \cite{BS} in {\sc Lepto} has been incorporated
using the scale $\min(\mu^2,W^2-\hat{s},\hat{s} + Q^2/2)$ where 
$\mu^2$ is set by LSTHFL(9) 
(default is $(m_{\perp Q} + m_{\perp \bar{Q}'})^2$).

The multiparton final state is then supplemented by the proton remnant.
In case of an incoming gluon, the remnant contains the three valence
quarks. The energy-momentum of the remnant is divided and a small
relative transverse momentum introduced when this system is split
\cite{TARGET,LEPTO} into a quark and a diquark. Together with the
produced heavy antiquark and quark, respectively, these spectator
partons form  two separate colour triplet strings. Before hadronising
these strings with {\sc Jetset} 7.4 \cite{JETSET} extra soft colour
interactions may be taken into account\footnote{This will give some
production of $q\bar{q}$ resonances for sufficiently small masses
of the $q\bar{q}$ system.
However, the treatment is rather crude and only $J/\psi$ and
$\Upsilon$ are produced for $c\bar{c}$ and $b\bar{b}$ respectively.} which is regulated by LST(34).
This results in a Monte Carlo model for heavy flavour production in
$ep$ collisions that generates complete events with the full
information on both the parton and the hadron level.

\section{Description of program components}
The program is written in FORTRAN 77 and consists of a set of
subroutines that form an add-on package to {\sc Lepto} 6.5 \cite{LEPTO}.
In addition, {\sc Jetset} 7.4 and {\sc Pythia} 5.7 \cite{JETSET} and 
the {\sc Divonne} program \cite{DIVON}, which are all available via
{\sc Cernlib}, are used.
The user has to ensure that the {\sc Aroma} routines are loaded
such that they replace some routines in {\sc Lepto} with their modified
versions, i.e. normally {\sc Aroma} should be loaded first.
All subroutine and common-block names start with HF
to avoid name clashes.
Exceptions are: AROMA and AROINI which the user calls;  DFUN and DVNOPT 
which replaces subroutines in {\sc Lepto} 6.5 and the block data ARODAT.

The heavy flavour generation is made by calling AROMA.
The parameters, cuts and switches for heavy flavour generation are in
arrays PARHFL, CUTHFL and LSTHFL of common HFDATA described below.

\subsection{Subroutines and functions}
The following routines are called by the user:

\noindent
SUBROUTINE {\bf AROINI}(LFILE,LEPIN,PLZ,PPZ,INTER)
\begin{defl}{123456789012}
\item[{\it Purpose:}]
to initialize the event generation procedure, see also \cite{LEPTO}.
\item[{\it Arguments:}]
\item[LFILE~:]  $=0$ normally when using {\sc Aroma}.
\item[LEPIN~:] type of lepton in {\sc Jetset} 7.4 code,
i.e. $11=e^-,\: -11=e^+$.
\item[PLZ, PPZ~:]
momentum (GeV/c) for incoming lepton and nucleon, respectively,
along the z-axis (if both non-zero, i.e.\ colliding beams, they must
have opposite signs).
\item[INTER~:]
type of interaction to be simulated.
\item[{\hfill =1:}]  electromagnetic (EM), i.e. $\gamma$ exchange.
\item[{\hfill =2:}]  weak charged current (CC), i.e. $W^{\pm}$ exchange.
\item[{\hfill =3:}]  weak neutral current, i.e. $Z^0$ exchange.
\item[{\hfill =4:}]  neutral current (NC), i.e. $\gamma /Z^0$ exchange.
\end{defl}

\noindent
SUBROUTINE {\bf AROMA}
\begin{defl}{123456789012}
\item[{\it Purpose:}]
to administer the generation of one event of the kind specified by the
last AROINI call.
\item[{\it Remark:}]
If the error flag LST(21) is non-zero then the event should
be rejected (should only occur rarely).
\end{defl}

\noindent
SUBROUTINE {\bf LFRAME}(IFRAME,IPHI)
\begin{defl}{123456789012}
\item[{\it Purpose:}]  to transform the event between different frames,
                       see \cite{LEPTO}.
\end{defl}

In the following list all subroutines (S) and functions (F) are briefly
described. The order is as they appear in the code and reflects the
flow in the program. Further details about their purpose and procedures
used are given by comments in the code.
\noindent
\begin{defl}{123456789012}
\item[{\it Routine}]  {\it Purpose}
\item[ARODAT \hfill ~~~]
     Block data to give default values to all switches and parameters.
\item[AROINI \hfill (S)] To initialize program, see above.
\item[HFINIT \hfill (S)]
             Initiate constants, calculate integrated cross section,
             set-up grid for event generation.
             Called by AROINI.
\item[AROMA \hfill (S)] To generate an event, see above.
\item[HFLGEN \hfill (S)]
                Administer heavy flavour generation based on exact
                matrix elements, called by AROMA. 
\item[DVNOPT \hfill (S)] Substitutes block data for {\sc Divonne}.
\item[DFUN   \hfill (F)] Integrand for {\sc Divonne}, calls HFFUNC.
\item[HFFUNC \hfill (F)]
            Differential cross section summed over flavours.
\item[HFVECT \hfill (S)]
             Transforms XX(1), $\ldots$, XX(4) to $y$, $Q^2$, $x_g$,
             $z$.
\item[HFCHCK \hfill (S)]
            Check if $y,Q^2,x_g$ and $z$ are within bounds for given
            flavours.
\item[HFTOTD \hfill (F)]
            Differential cross section for given flavours.
\item[HFVCT1 \hfill (S)]
  Variable mapping for $\gamma$-exchange, called by HFVECT.
\item[HFVCT2 \hfill (S)]
  Variable mapping for $W$- and $Z$-exchange, called by HFVECT.
\item[HFMEPS~(S)] Modified version of LMEPS in {\sc Lepto}.
\item[HFSCAL~(S)] Modified version of LSCALE in {\sc Lepto}.
\item[HFSSPA~(S)] Modified version of LYSSPA in {\sc Lepto}.
\item[HFANGL~(F)] Modified version of ULANGL in {\sc Jetset}.
\item[HFDBRB~(S)] Modified version of LUROBO in {\sc Jetset}.
\end{defl}

\noindent
The following subroutines in {\sc Divonne} 4 are used:
PARTN, INTGRL, RANGEN.

\subsection{Common blocks}

The common blocks intended for communication with the program are
HFDATA and LEPTOU. Common LUJETS in {\sc Jetset} 7.4 \cite{JETSET} 
is used to store the event record and is therefore essential.
All variables are given sensible default values in the
block data ARODAT, as shown by (D=\ldots ) below.
Overwriting the default values should be made before calling AROINI.

\newpage
\noindent
COMMON /{\bf HFDATA}/ PARHFL(10),CUTHFL(8),LSTHFL(10)
\begin{defl}{123456789012345}
\item[{\it Purpose:}]
     contains input switches and input parameters to specify physics,
     kinematic cuts and numerical procedures, as well as output flags.
     Note that PARHFL and CUTHFL are in double precision.
\item[PARHFL(1)~:]
                total cross section in $pb$ for heavy flavour production.
                Calculated in the initialization taking the hard cuts 
                in CUTHFL into account.
\item[PARHFL(2)~:]
                (D=0.2) precision parameter SPRDMX for call to PARTN in
                 the {\sc Divonne} program;
                smaller values give finer partitioning.
\item[PARHFL(3)~:]
                (D=1000.) parameter NPT for call to INTGRL in {\sc Divonne}.
                Specifies maximum function evaluations in each subregion
                for integrating
                total heavy flavour cross section.
\item[PARHFL(4)~:]
(D=0. GeV$^2$) defines the minimum $p_\bot^2$ of the quarks
in the boson-gluon cms, must be used when producing light quarks.
\item[PARHFL(5)~:] fraction of phase space points generated within
the `hard' cuts given in CUTHFL (see below) that are accepted by the `soft' 
cuts in CUT (see common LEPTOU in \cite{LEPTO}) when generating the
complete events. 
\item[PARHFL(6)~:] cross-section in $pb$ for the generated event sample. 
Given by the product of PARHFL(1) and PARHFL(5), thus taking  
both hard and soft cuts into account. To be used at the end of event 
generation (updated for each event) to get absolute cross section normalisation
of generated event sample.
\item[PARHFL(7-10)~:] not used
\item[~~ ]
\item[CUTHFL(1)~:] $y_{\min}$ (D=0.)
\item[CUTHFL(2)~:] $y_{\max}$ (D=1.)
\item[CUTHFL(3)~:] $Q^2_{\min}$ (D=0.) (GeV$^2$), can be set to zero when
     only heavy flavour production is switched on (cf. LSTHFL(1)).
\item[CUTHFL(4)~:] $Q^2_{\max}$ (D=1.D9) (GeV$^2$)
\item[CUTHFL(5)~:] $x_{g,\min}$ (D=0.)
\item[CUTHFL(6)~:] $x_{g,\max}$ (D=0.9999999)
\item[CUTHFL(7)~:] $z_{\min}$ (D=0.)
\item[CUTHFL(8)~:] $z_{\max}$ (D=1.)
\item[\hfill {\it Remark:}] These `hard' cuts are on the basic variables in 
     the heavy flavour matrix element and can therefore be efficiently 
     applied to limit the region of phase space. (Kinematic phase space limits
     are explicitly calculated and applied if stronger than these cuts.)  
     In contrast, the `soft'
     cuts in CUT (common LEPTOU in \cite{LEPTO}) are used to reject 
     chosen points or partly generated events, resulting in less 
     efficiently applied cuts. See PARHFL(5), PARHFL(6) above. 
\item[~~ ]
\item[LSTHFL(1)~:]     (D=2) regulates heavy flavour generation.
\item[{\hfill =0:}]  DIS events, no heavy flavour production.
\item[{\hfill =1:}]  DIS events and heavy flavour production.
\item[{\hfill =2:}]  only heavy flavour production.
\item[\hfill {\it Remark:}]
    When `normal' {\sc Lepto} events and heavy flavour events are mixed in the
    same run, the total cross section is then the sum of the normal
    one (PARL(23) or PARL(24)) and the heavy flavour one (PARHFL(6)).
    To avoid double counting precautions have been taken in AROINI by setting
    switches such that the same flavours should not be
    produced in other processes; i.e.\ boson-gluon fusion process with massless
    matrix elements in {\sc Lepto} (LST(13)), through parton showers 
    (MSTJ(45), IPY(8)) and via sea quark density parametrizations (LST(12)).
\item[LSTHFL(2), LSTHFL(3)~:]
               (D=4,5) lightest and heaviest flavour to be produced in
               NC heavy flavour production,
               can be set between 1 and 6 for $d,u,s,c,b,t$.
\item[LSTHFL(4)~:] (D=4) specifies process simulated
\item[{\hfill =1:}]  pure $\gamma$ contribution.
\item[{\hfill =2:}]  $W$ exchange.
\item[{\hfill =3:}]  pure $Z$ contribution.
\item[{\hfill =4:}]  full NC contribution ($\gamma + \gamma /Z + Z$).
\item[LSTHFL(5)~:] internal, $>0$ for $e^+$, $<0$ for $e^-$.
\item[LSTHFL(6)~:]
                (D=11) regulates which flavour combinations to be
                included in CC interactions.
                Should be set to $I_{cs} + 10 I_{cb} + 100 I_{ts}
                + 1000 I_{tb}$, where
                $I_{cs}$ etc.\ is $0$ to exclude ($1$ to include)
                $c\bar{s}$ (or $\bar{c}s$) etc.
\item[LSTHFL(7)~:]  (D=4) regulates the scale  in the final state
                parton showers in $Q\bar{Q}$ system.
                Value is multiplied by PYPAR(25).
\item[{\hfill =1:}] $Q^2$,
\item[{\hfill =2:}] $W^2$,
\item[{\hfill =3:}] $\hat{s}$, {\it i.e.} 
                    same treatment as for $q\bar{q}$-event in {\sc Lepto}
\item[{\hfill =4:}] $(m_{\bot Q} + m_{\bot \bar{Q}})^2$,
\item[{\hfill =5:}] $m_{\bot Q}^2 + m_{\bot \bar{Q}}^2$,
          where $m_{\bot Q}$ is the transverse mass of the heavy quark.
\item[LSTHFL(8)~:] (D=2) simulation of QCD effects in hadronic final
     state, see \cite{LEPTO}.
\item[{\hfill =1:}]
     no parton showers
\item[{\hfill =2:}]
     both initial and final state parton showers
\item[{\hfill =3:}]
     initial state parton shower
\item[{\hfill =4:}]
     final state parton shower
\item[LSTHFL(9)~:] (D=4) choice of scale, i.e. maximum parton
     offshellness, in QCD initial state cascade.
     Value is multiplied by PYPAR(26).
\item[{\hfill =1:}] $Q^2$,
\item[{\hfill =2:}] $W^2$,
\item[{\hfill =3:}] $\hat{s}$,
\item[{\hfill =4:}] $(m_{\bot Q} + m_{\bot \bar{Q}})^2$,
\item[{\hfill =5:}] $m_{\bot Q}^2 + m_{\bot \bar{Q}}^2$,
\item[{\hfill =6:}] Same treatment as for $q\bar{q}$-event in {\sc Lepto}.
\item[\hfill {\it Remark:}]
     For the initial state shower the scale used is
     $\min(\mu^2,W^2-\hat{s},\hat{s}+Q^2/2)$ where $\mu^2$ is the
     scale chosen above ($Q^2$ should not be used for photoproduction).
\end{defl}

\noindent
COMMON /{\bf LEPTOU}/  CUT(14),LST(40),PARL(30),X,Y,W2,Q2,U
\begin{defl}{123456789012}
\item[{\it Purpose:}]
     contains input switches and input parameters to specify physics,
     soft kinematic cuts and numerical procedures, as well as output flags
     and output variables.
\item[{\it Remark:}] only changes for {\sc Aroma} are listed
     here, otherwise see \cite{LEPTO}.
\item[LST(8)~:]  set by AROINI corresponding to LSTHFL(8).
\item[LST(17)=1:]  not usable with {\sc Aroma}.
\item[LST(24)=5:]  current event is heavy quark event from {\sc Aroma}.
\item[LST(35)~:]  there is no special sea quark treatment in the
                  initial cascade in {\sc Aroma}.
\end{defl}

\noindent
COMMON /{\bf HFLVAR}/ SHAT,ETA,Z,PHIQQ,PT2Q,TM2Q1,TM2Q2,ZJ,IFLQ,IFLQB,
    \& PHIFL(0:9,0:2),DBETAZ
\begin{defl}{123456789012}
\item[{\it Purpose:}] Variables at the parton level, values obtained
                    from simulation of cross section. Note that DBETAZ is
                    of type double precision.
\item[SHAT,ETA,Z,PHIQQ~:] variables $\hat{s}$, $x_g$, $z$ and $\Phi$,
                          see section 1.
\item[PT2Q~~~:] $p^2_\bot$ for produced $Q$ (and $\bar{Q}$)
                in the boson-gluon cms, see section 1.
\item[TM2Q1,TM2Q2~:] $m^2_{\bot}$ for produced $Q$ and $\bar{Q}$
                in the boson-gluon cms.
\end{defl}

\noindent
COMMON /{\bf LFLMIX}/ CABIBO(4,4)
\begin{defl}{123456789012}
\item[{\it Purpose:}]
     Contains the Cabbibo-Kobayashi-Maskawa matrix elements squared
     for flavour mixing, see \cite{LEPTO}.
\end{defl}

\noindent
COMMON /{\bf LUDAT2}/ ...PMAS(500,4)...
\begin{defl}{123456789012}
\item[{\it Purpose:}] Various parameters in {\sc Jetset} \cite{JETSET}.
\item[PMAS(IFL,1)~:]
     mass for quark IFL=1--6 for $d,u,s,c,b,t$ used in the BGF matrix element
\end{defl}

\noindent
COMMON /{\bf LUJETS}/ N,K(4000,5),P(4000,5),V(4000,5)
\begin{defl}{123456789012}
\item[{\it Purpose:}]
Contains the record of the generated event, see \cite{JETSET}.
\item[{\it Remark:}]
The event record contains the full development of the event including
a few extra, but inactive lines used for internal testing purposes.
\end{defl}

\noindent
The following common-blocks are mainly for internal use: 
HFBL1, HFBL2, HFJETS, HFJRDM, HFMACH, HFSNCO.\\
{\sc Divonne} common-blocks used: BNDLMT, SAMPLE, PRINT, QUADRE.

\section{Usage and Availability}
{\sc Aroma} 2.2 should be loaded together with {\sc Divonne} 
\cite{DIVON}, 
{\sc Lepto 6.5} \cite{LEPTO}, 
{\sc Jetset 7.4} and {\sc Pythia 5.7} \cite{JETSET}. 
The program is a slave system, which the user must call from
his own steering program.
Information about the program, the source code and a demonstration job 
can be obtained from the authors or via the WWW on the {\sc Aroma} 
homepage,
\texttt{http://www3.tsl.uu.se/thep/aroma/}.


\end{document}